\documentclass[]{PoS}

\def\degr{\hbox{$^\circ$}}
\def\arcmin{\hbox{$^\prime$}}
\def\farcs{\hbox{$.\!\!^{\prime\prime}$}}
\def\fsec{\hbox{$.\!\!^{\rm s}$}}

\title{EVN observations of an OH maser burst in OH17.7$-$2.0}

\ShortTitle{EVN observations of an OH maser burst in OH17.7$-$2.0}

\author{\speaker{M. Szymczak} and A. Bartkiewicz \\
        Toru\'n Centre for Astronomy, Nicolaus Copernicus University, Poland\\
        E-mail: \email{msz@astro.uni.torun.pl, annan@astro.uni.torun.pl}}

\author{E. G\'erard \\
        Observatoire de Paris, France\\
        E-mail: \email{Eric.Gerard@obspm.fr}}

\abstract{We have observed the OH 1612-MHz maser emission 
towards the proto-planetary nebula candidate OH17.7$-$2.0 
that underwent a very strong and unusual outburst in 2003.
Phase-referencing data were obtained with the EVN 
in order to localize the outburst and to examine its
possible causes. The majority of the emission comes from an 
incomplete spherical shell with inner and outer radii of 220 
and 850\,mas, respectively. There is a strong evidence for
maser components that arise due to the interaction of a jet-like
post-AGB outflow with the remnant outer AGB shell.
The most prominent signature of such an interaction is the strongly 
bursting polarized emission near 73.3\,km\,s$^{-1}$ 
coming from two unresolved components of brightness 
temperature up to $10^{11}$\,K located at the edge of
the biconal region 2500\,AU from the central star.
It is remarkable that this OH biconal region is well-aligned 
with the polar outflow inferred from the near-infrared image.
\
}

\FullConference{The 8th European VLBI Network Symposium \\
		September 26-29 2006\\
		Toru\'n, Poland}

\begin{document}

\section{Background and motivation}
There are several lines of evidence that OH17.7$-$2.0 is an
oxygen-rich post-AGB object. Its central star has a spectral 
type earlier than $\sim$K5 with an effective temperature 
$\sim$4000 $-$ 10000\,K (\cite{lebertre89}). This is a dusty
object with far-infrared colours consistent with a detached shell 
and mass-loss rate of $\sim4\times 10^{-5}$M$_{\odot}$yr$^{-1}$
(\cite{vanderveen95}). At 2.2\,$\mu$m it appears as a prolate 
spheroid with the polar axis at a position angle of 20$^{\circ}$
(\cite{bains03}). Non-variability or only slight variations at 
near-infrared wavelengths (\cite{lebertre89}), small amplitude
OH flux variations at 1612\,MHz (\cite{herman85}), disappearance
of the H$_2$O maser (\cite{engels02}) and non-detection of SiO masers
(\cite{nyman98}) are consistent with a post-AGB status of 
OH17.7$-$2.0. 

In 2003, the OH 1612-MHz maser line profile underwent a very 
unusual change; the intensity of the red-shifted emission at 
velocity $\sim$73\,km\,s$^{-1}$ increased nearly threefold over
a period of $\sim$430\,days (\cite{szymczak05}). The flaring 
feature exhibits strong (up to 80\%) circular polarization
and considerable ($\sim$15\%) linear polarization. 
It is remarkable that adjacent red-shifted features do not show
variations within an absolute flux density uncertainty less than
10\%. Furthermore, the entire blue-shifted part of the 1612-MHz 
spectrum  does not show any variations either. 
Finally, no changes were detected in the integrated flux 
densities of the 1665 and 1667-MHz maser lines (\cite{szymczak05}).
 
The outburst event in OH17.7$-$2.0 is spectacular and unique 
when compared to flare phenomena in circumstellar OH masers reported 
in the literature. Almost all eruptive changes in OH maser emission
observed so far took place in Mira-type variables and are
characterized by global changes with time scales of months and years
in the OH profiles of the 1612, 1665 and 1667-MHz transitions when 
present (\cite{jewell81}, \cite{etoka97}), usually weakly associated 
with changes in the optical and/or infrared.

In this paper, we report the preliminary results of follow-up VLBI 
observations taken in order to localize the burst in the shell and 
to determine the properties of the active regions. This should allow us 
to constrain possible causes of the outburst (\cite{szymczak05}) 
and to understand the evolutionary status of the central star.

\section{Observations and reduction}
The OH 1612-MHz observations were carried out on 2005 March 3 
using eight EVN telescopes: Cambridge, Effelsberg, Hartebeesthoek,
Jodrell Bank, Medicina, Onsala and Toru\'n. 
The observations consisted of six 38-min. scans on the target 
interleaved with 6-min. scans on the phase-calibrator source 
J1733$-$1304. One-hour scan of the continuum source J2253+1608
was also made for the purpose of bandpass and amplitude
calibration. Dual circular polarization was recorded using the
MarkV system with a spectral bandwidth of 0.5\,MHz. 
For OH17.7$-$2.0, this translates to LSR velocities from 15 to 
108\,km\,s$^{-1}$. The data were correlated with the JIVE correlator 
using 1024 channels, yielding a spectral resolution of 0.09\,km\,s$^{-1}$. 
Correlated data were calibrated and reduced using AIPS package, 
following the standard procedures for spectral-line 
VLBI observations. The resulting synthesized beam size was 
64$\times$20\,mas at a position angle 9$^{\circ}$.
An area of $2\times2$~arcsec$^2$ was searched for maser emission
above 5$\sigma$ limit over the entire band.
For single circular polarization, the rms noise in the final 
emission-free channel maps was $\sim$9\,mJy.
The relative positions of the maser components were determined
with 0.5\,mas accuracy.

\section{Results}
OH 1612-MHz maser emission was detected at velocities between
47.9 and 75.0\,km\,s$^{-1}$; this is a slightly narrower range than
that reported in MERLIN observations (\cite{bains03}). 

\begin{figure}
\centering
\includegraphics[width=.6\textwidth]{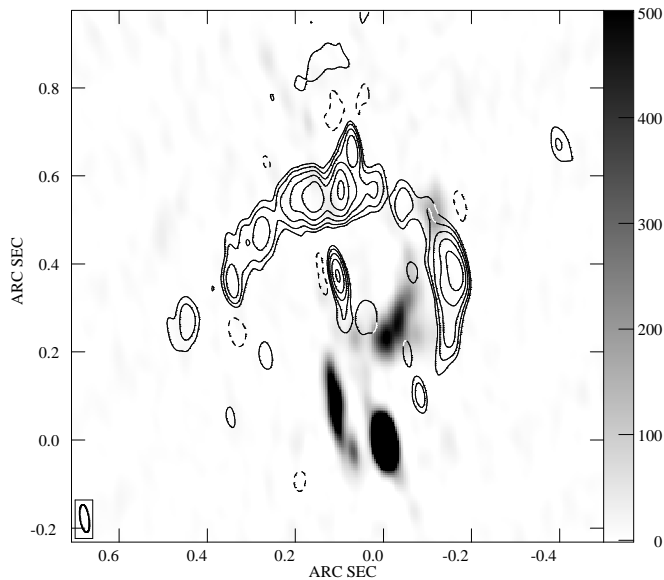}
\caption{Integrated intensity map of the OH 1612-MHz maser emission 
  of OH17.7$-$2.0 for left-hand circular (LHC) polarization. 
  The contours plotted at intervals of
  (-1, 1, 2, 4, 8, 16)$\times$20\,mJy\,beam$^{-1}$\,m\,s$^{-1}$ 
  show the blue-shifted emission ($V<$61.6\,km\,s$^{-1}$). 
  The red-shifted emission ($V>$61.6\,km\,s$^{-1}$) is shown in grey 
  that scales from 0 to 500\,mJy\,beam$^{-1}$\,m\,s$^{-1}$. 
  Note that, for better presentation, the grey scale is cut at 
  500\,mJy\,beam$^{-1}$\,m\,s$^{-1}$ level and the peak intensity of 
  the southern unresolved component is
  9931\,mJy\,beam$^{-1}$\,m\,s$^{-1}$. 
  The map origin is at RA(J2000)=18$^{\rm h}$30$^{\rm m}$30\fsec695,
  Dec(J2000)=$-$14\degr28\arcmin56\farcs82.
  The ellipse in the bottom left corner of the map represents 
  the restoring beam.}
\label{fig1}
\end{figure}

Fig.\,1 shows the LHC map of line emission integrated over all channels. 
Most of the blue-shifted emission originates from a ring-like
structure which resembles a symmetric shell of about 510~mas
diameter. No emission is seen from the S
and the SE parts of the shell. Compact and unresolved 
emission in the velocity range of 48.6$-$49.6\,km\,s$^{-1}$ is found at 
a position which coincides, within the synthesized beam, with the
position of the central star inferred from the model of expanding
spherical shell. Thus, this compact emission is very likely the amplified
stellar image. Weak and diffuse maser components at blue-shifted
velocities higher than 48.5\,km\,s$^{-1}$ are seen $\sim$380\,mas 
to the SE and $\sim$600\,mas to the NW of that compact emission 
(the amplified stellar image), at a position angle of 
$\sim$160$^{\circ}$. The red-shifted diffuse emission is also detected at
roughly the same position angle but is weakly visible in Fig.\,1 due 
to dynamic range limitations. Weak and scattered emission likely comes
from an equatorial region. The red-shifted emission is dominated by
two very strong (>150\,Jy\,beam$^{-1}$) unresolved components at
velocities $\sim$73\,km\,s$^{-1}$ and at a projected distance of
$\sim$400\,mas south of the central star. The brightness temperature
of these bursting components is up to 10$^{11}$\,K.  The rest of the
red-shifted emission follows a ring-like distribution with the SW part
clearly seen. Clearly, there is a lack of red-shifted emission from
the northern side of the shell. 

\section{Interpretation}
In general, the masers are located within a remnant spherical shell
(Fig. 2). However, there is evidence for several major departures from 
the simple shell model. (1) There is an offset of $\sim$300\,mas
between the two emission peaks at extreme velocities along an axis 
at position angle of 8.4$^{\circ}$. (2) Along the same position angle, 
a biconal region can be seen; the emission appears to come from the sides 
of the cones as projected on the sky and likely represents
the interaction of a faster post-AGB wind or jet with the outer AGB wind.
In this case, the boundary appears to be thin and tangential emission
may dominate. The opening angle of the cone is less than 15$^{\circ}$
while the inclination angle between the equatorial plane and the line
of sight is $\sim30^{\circ}$. The red-shifted bubble is at 
projected distance of $\sim400$\,mas. (3) Low-brightness emission at
middle radial velocities scattered along a plane roughly orthogonal to
the axis of the cones is likely due to the emission from the denser
equatorial region or torus-like structure. Such a region can be
produced during a period of enhanced mass loss at the end of the AGB 
phase. The OH maser emission from this region, as predicted by a spherical 
shell model, should be weak due to short amplification path length. 

It is remarkable that the position angle of the axis of biconal region 
seen in the OH 1612-MHz maser distribution ($\sim10^{\circ}$) is 
quite consistent with the position angle of the major axis of the
infrared nebulosity of size 2.5\,arcsec ($20^{\circ}$, \cite{bains03}).
This confirms that OH17.7$-$2.0 recently started a bipolar outflow.

The position-velocity diagram (Fig. 3) implies that a spherical 
velocity field for the bulk of the 1612-MHz emission, in the first 
instance, satisfactorily fits all the data. The position of the central star 
is $\Delta\alpha$=74\,mas, $\Delta\delta$=396\,mas relative to the map
origin (Fig. 1) and coincides within the beam with the position of 
the compact blue-shifted emission which is interpreted as the
amplified stellar image. 
The best-fitted systemic and expansion velocities are 
61.6\,km\,s$^{-1}$ and 13.8\,km\,s$^{-1}$, respectively.   
The inner radius is 220\,mas while the outer radius is up to 850\,mas.
Although the envelope is sparsely filled with the maser components,
Fig. 3 shows evidence for multiple shells which may result from
periodic enhancements of mass loss rate within a range of 300-800
years, for the assumed distance of 3.4\,kpc. It appears that the maser
outburst occurred in a shell of radius of 740\,mas, i.e., 2500\,AU from
the central star. For the first time, the present data provide evidence  
that OH17.7$-$2.0 has a bipolar morphology of the 1612-MHz maser
emission and the active region is confined to the surface of bubble.
We argue that one of most plausible causes of the burst is an
interaction of a jet-like outflow with the remnant outer AGB shell.
If the disappearance of the H$_2$O maser in 1990 (\cite{engels02}) is 
casually related to the 1612-MHz OH outburst in 2003 and assuming
that the H$_2$O maser arose in an inner AGB shell ($\sim$50\,AU), 
the speed of the jet-like outflow should be $\sim$800\,km\,s$^{-1}$. 
The presence of such outflows is documented in many proto-planetary 
nebulae, but their effect on the onset of the OH maser emission is 
not well understood (\cite{szymczak05}). Our data suggest that 
the outburst emerges from a narrow interface between the surface 
of the bubble and dense clumps of the detached AGB shell where the 
magnetic field must play a role in the maser amplification since only 
one sense of circular polarization is seen.

\begin{figure}
\centering
\includegraphics[width=.6\textwidth]{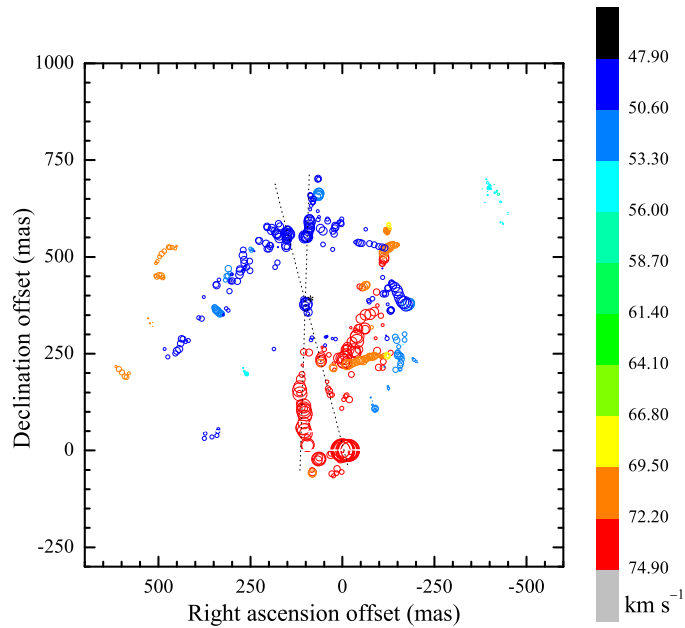}
\caption{The overall distribution of all maser components
  brighter than 10$\sigma$ ($>$100\,mJy\,beam$^{-1}$). The origin of
  diagram is the position of the brightest component. The size of the
  symbols is proportional to the logarithm of brightness. 
  The colour corresponds to the velocity scale shown by a vertical bar.
  The dashed lines outline the polar cavities where the maser emission
  is quenched while it is present at the biconal surface.}
\label{fig2}
\end{figure}

\begin{figure}
\centering
\includegraphics[width=.6\textwidth]{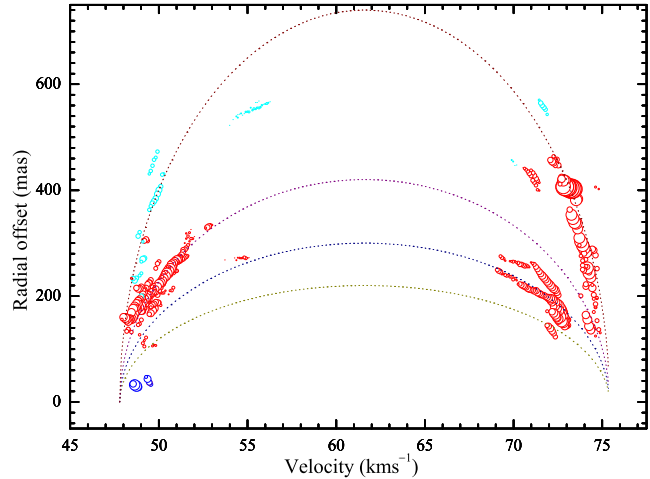}
\caption{Radial offset of the maser components from the best fitted
  stellar position as a function of the radial velocity. The circle
  size is proportional to the logarithm of the component brightness.
  Colours distinguish the components from the shells (red), the equatorial 
  region (cyan) and amplification of stellar photons (blue). 
  The dashed curves correspond to the models of spherical shells of
  radii 220, 300, 420 and 740\,mas assuming a constant outflow velocity
  of 13.8\,km\,s$^{-1}$.}
\label{fig3}
\end{figure}

MERLIN observations of OH17.7$-$2.0 revealed the southern outlying
components (\cite{bains03}) which are interpreted as the result of the
interaction of a fast post-AGB wind with the remnant outer AGB wind.
In this scenario, the OH masers also trace the radial outflow away from
the central star along the surface of the cones; the velocities of
these maser components increase linearly with the distance from the star 
(\cite{zijlstra01}).
The presence of such OH components was confirmed in some post-AGB
objects with broad ($>$50\,km\,s$^{-1}$) OH profiles
(\cite{zijlstra01}). We failed to find a similar phenomenon in our
target. This suggests that OH17.7$-$2.0 is in an early post-AGB
evolution phase.

\section{Concluding remarks}
The most striking result from the first-epoch EVN observations 
is the identification of an active region where the 1612-MHz outburst 
occurred. This finding strongly constrains a suite of possible causes
of the outburst (\cite{szymczak05}). It appears that one of the most
probable causes is the interaction of a jet-like outflow from the
post-AGB star with dense clumps of the remnant outer AGB shell. 
This confirms that OH17.7$-$2.0 is in a transitional, short-lasting
phase from AGB to PN. Multi-epoch VLBI observations would allow us 
to trace the changes in the structure of OH maser emission and 
to explain a possible role of the magnetic field in the shaping
of the bipolar outflow. It seems that this target is ideal
for optical and infrared observations in order to obtain a
complete picture of the changes occurring in the whole envelope.

\end{document}